\documentclass{mn2e}
\usepackage{epsfig,natbib2,natbibmnfix}
\usepackage{astrojournals}
\usepackage{amssymb}
\usepackage{multirow}
\usepackage{multicol}
\usepackage[varg]{txfonts}
\usepackage{color}
\usepackage{url}

\newcommand{\fref}[1]{Fig.~\ref{#1}}

\newcommand{\cref}[1]{Chapter~\ref{#1}}
\newcommand{\sref}[1]{Section~\ref{#1}}

\def \msun{\rm M_{\odot}}

\begin{document}
\title[Rapid accretion]{Rapid AGN accretion from counter--rotating discs}
\author[C.~J.~Nixon, A.~R.~King and D.~J.~Price ] {
\parbox{5in}{Christopher~J.~Nixon$^{1\star}$, Andrew~R.~King$^1$ and
  Daniel~J.~Price$^2$}
\vspace{0.1in} \\ $^1$ Theoretical Astrophysics Group, University of
Leicester, Leicester LE1 7RH UK \\ $^2$ Monash Centre for Astrophysics (MoCA),
 School of Mathematical Sciences, Monash University, Vic.
3800, Australia}
\maketitle

\begin{abstract}
Accretion in the nuclei of active galaxies may occur chaotically.  This can
produce accretion discs which are counter--rotating or strongly misaligned
with respect to the spin of the central supermassive black hole (SMBH), or the
axis of a close SMBH binary.  Accordingly we consider the cancellation of
angular momentum in accretion discs with a significant change of plane (tilt)
between inner and outer parts. We estimate analytically the maximum accretion
rate through such discs and compare this with the results of Smoothed Particle
Hydrodynamics (SPH) simulations. These suggest that accretion rates on to
supermassive black holes may be larger by factors $\gtrsim 100$ if the disc is
internally tilted in this way rather than planar. This offers a natural way of
driving the rapid growth of supermassive black holes, and the coalescence of
SMBH binaries.
\end{abstract}

\begin{keywords}
{accretion, accretion discs -- black hole physics -- galaxies:
  formation -- galaxies: active -- cosmology: theory} 
\end{keywords}

\footnotetext[1]{E-mail: chris.nixon@astro.le.ac.uk}
\section{Introduction}
\label{intro}
Astronomers still do not know how the supermassive black holes (SMBH)
in galaxy centres accrete gas and grow. The problem lies in the very
small angular momentum this gas must have in order to accrete on a
reasonable timescale. For example, the viscous time of a standard
accretion disc of $\sim 1$~pc radius accreting on to a SMBH of
$10^8\msun$ approaches a Hubble time. A natural remedy for this is to
assume that on small scales near the hole, accretion occurs in
episodes whose angular momenta are randomly oriented with respect to
each other. This chaotic picture predicts that the spin of the hole
remains low, allowing rapid mass growth provided that there is an
adequate mass supply (\citealt{KP2006}; \citealt{KP2007};
\citealt{Kingetal2008}; \citealt{Fanidakisetal2011}). A characteristic
feature of this type of accretion is that the disc flow on to the SMBH
is often retrograde \citep{Kingetal2005} with respect to the hole
spin, or to the rotation of an SMBH binary resulting from a galaxy
merger.

In a recent paper \citep{Nixonetal2011a} we showed that a retrograde
coplanar external gas disc can be markedly more efficient in shrinking
an SMBH binary than a prograde one. A retrograde disc can cancel the
binary orbital angular momentum directly, rather than being slowed by
the resonances which always arise in a prograde circumbinary disc. It
is natural to ask if this angular momentum cancellation can work if we
replace the SMBH binary with a pre--existing gas disc surrounding a
single SMBH. If so, this may be a mechanism promoting much faster
delivery of gas on to the black hole. This situation may arise
naturally from misaligned accretion events, for example where an
accretion event forms a disc around the black hole, and a second event
forms an outer disc which is misaligned with respect to the first
one. Similar cases occur as a misaligned accretion disc attempts to
reach an axisymmetric (co-- {\it or} counter--aligned) configuration
around a spinning black hole (\citealt{Kingetal2005};
\citealt{LP2006}), and also during the closely related co-- or
counter--alignment of an accretion disc around a central binary
\citep*{Nixonetal2011b}.

In all these cases integrating the full dynamics of the accretion disc
is complex and time--consuming because the component discs can warp
before interacting. To study the mixing of distinct or opposing disc
angular momenta we investigate a simple form of this kind of
accretion. We assume an accretion disc with initially distinct
planes in its inner and outer parts. We let these two parts interact
viscously and determine how much the accretion is enhanced compared
with a disc lying in a single plane. In \sref{analytical} we give
simple analytical estimates of this. In \sref{sims} we perform global
3D hydrodynamical simulations to examine the disc's dynamical
behaviour in the light of these estimates. We interpret our results in
\sref{discussion}.

\section{Counter--rotating Discs}
\label{analytical}
We consider an inner and outer disc counter--rotating with respect to
each other but not coplanar, that is, the angle $\theta$ between the
disc angular momentum vectors obeys $\pi/2 < \theta < \pi $.

If the discs were in perfect contact the gas would share opposed
angular momenta from almost the same radii. This may be unphysical (or
imply a very rapid evolution away from the initial state) since the
discs may attempt to open a gap of some size, as we discuss later. For
generality we choose an arbitrary gap between the discs of size $a$.

Now we assume that the gas efficiently shares angular momentum across
this gap and that an equal mass of gas from each disc takes part in
the interaction. The resultant angular momentum of the interacting gas
is
\begin{eqnarray}
\lefteqn{L_{\rm new} =\frac{1}{2} \left[L\left(R_{\rm gap}+a\right)
  + L\left(R_{\rm gap}\right)\right]} \nonumber
\\ \lefteqn{\ \ \ \ \ \ \ \ \approx \frac{1}{2}\sqrt{GMR_{\rm gap}}\left(
  1 + \frac{1}{2}\frac{a}{R_{\rm gap}} + \cos\theta \right)}
\label{Phi2}
\end{eqnarray}
to first order in $a/R_{\rm gap}$, where $R_{\rm gap}$ is the outer
edge of the inner disc (so that $R_{\rm gap} + a$ is the inner edge of
the outer disc), $G$ is the gravitational constant, $M$ is the mass
of the central object, and we have assumed a Keplerian potential
for the gas.

The circularisation radius for this interacting gas is
\begin{equation}
  R_{\rm circ} = \frac{1}{4}R_{\rm gap}\left( 1 + \cos\theta +
  \frac{1}{2}\frac{a}{R_{\rm gap}} \right)^{2}
\end{equation}
We expect the size of the gap between the two discs to be roughly the
scale of the epicyclic motions in the gas disc as this is where fluid
orbits cross and interact. This implies a gap size $\sim H(R_{\rm
  gap})$, where $H$ is the semi--thickness of the disc, so that
\begin{equation}
  R_{\rm circ} = \frac{1}{4}\left( 1 + \cos\theta + \frac{1}{2}\frac{H}{R}
  \right)^{2}R_{\rm gap}.
\label{rcirc1}
\end{equation}
For typical AGN disc parameters, $H/R \sim 10^{-3}$, and
there are two distinct cases. For $\theta$ close to $\pi$, we can have
$1+\cos\theta \ll H/R$, so that
\begin{equation}
  R_{\rm circ} \approx \frac{1}{16}\left(H/R \right)^{2}R_{\rm gap}.
\label{rcirc2}
\end{equation}
This suggests that the gas orbits would decrease by $\sim 7$ orders of
magnitude in radius which implies direct accretion of the gas as long
as its path to the hole is clear.

If however $\theta$ is not close to $\pi$ we have $1+\cos\theta \gg
H/R$, and
\begin{equation}
  R_{\rm circ} \approx \frac{1}{4}\left( 1 + \cos\theta \right)^{2}R_{\rm gap}.
\label{rcirc3}
\end{equation}
Then the gas orbit decreases least when $\theta \sim \pi/2$, where
$R_{\rm circ} \sim \frac{1}{4}R_{\rm gap}$. The orbit scale
(\ref{rcirc1}) varies smoothly between these two extreme values.

The accretion rate through a disc is approximately
\begin{equation}
  \dot M \sim \frac{M_{\rm disc}}{\tau_{\rm visc}}
\end{equation}
where $M_{\rm disc}$ is the disc mass and $\tau_{\rm visc}$ is the
viscous timescale
\begin{equation}
  \tau_{\rm visc} \sim \frac{R^{2}}{\nu}
\end{equation}
where $R$ is a characteristic radius for the disc and $\nu$ is a
measure of the disc viscosity. Thus for constant
viscosity\footnote{This is only approximate as one expects $\nu$ to
  increase with radius.} the accretion rate is $\propto R^{-2}$. This
suggests that for $\theta > \pi/2 $, where the gas orbits
decrease by a factor $> 4$, there is potential for the accretion
rate to increase by a factor $> 16$. For large $\theta \approx \pi$
the cancellation can lead to direct accretion of gas on a dynamical
timescale. This allows accretion on a local disc filling
timescale at the radius of the gap. This timescale is much shorter
than the viscous timescale as we only need move gas to $R_{\rm gap}$
and not right on to the hole. Note that the rate at which mass can be
supplied to $R_{\rm gap}$ is clearly an upper bound on the possible
accretion rate from this process.

If the gas does not accrete directly then the viscous timescale is shortened
to
\begin{equation}
\label{tvisc}
\tau_{\rm visc,~circ} \sim \left(R_{\rm circ}/R_{\rm
  gap}\right)^2\tau_{\rm visc,~gap}.
\end{equation}

Although these estimates are suggestive, we caution that we have
assumed that the gas shares angular momentum efficiently, and that an
equal mass of gas interacts from each disc. In reality neither of
these simplifications may be valid. Accordingly we use Smoothed
Particle Hydrodynamics (SPH) to simulate the interactions between the
discs.
\section{Simulations}
\label{sims}
\subsection{Code Setup}
We use \textsc{phantom}, a low-memory, highly efficient SPH code optimised for the study of non-self-gravitating problems. This code has performed well in
related simulations. For example \citet{LP2010} performed simulations
of warped accretion discs and found excellent agreement with the
analytical work of \citet{Ogilvie1999} on the nature of the internal
accretion disc torques in response to warping.

 The implementation of accretion disc $\alpha-$viscosity \citep{SS1973} in \textsc{phantom} is described in \citet{LP2010}. Specifically, we use the `artificial viscosity for a disc' described in Sec. 3.2.3 of \citet{LP2010}, similar to earlier SPH accretion disc calculations \citep[e.g.][]{Murray1996}. The main differences compared to standard SPH artificial viscosity are that the disc viscosity is applied to both approaching and receding particles and that no switches are used. Our implementation also differs slightly from \citet{LP2010} in that we retain the $\beta^{\rm AV}$ term in the signal velocity in order to prevent particle interpenetration from occurring in high-velocity stream-disc collisions. The disc viscosity in \textsc{phantom} was extensively calibrated against a 1D thin $\alpha-$disc evolution in \citet{LP2010} (c.f. Fig.~4 in that paper) and the disc scale heights employed here are similar. As the exact value of $\alpha$ is unimportant given the dynamical nature of the simulations, we simply use $\alpha^{\rm AV} = 1$ and $\beta^{\rm AV} = 2$ which at the employed resolution (see below) corresponds to a physical viscosity $\alpha_{\rm SS} \approx 0.02$--$0.06$.

For these simulations we model the gravity of the accretor with a Newtonian
point mass potential, adopting an accretion radius of 0.1 in code units. We
start with a flat disc of gas, composed of 10 million SPH particles, in
hydrostatic equilibrium between $1.0$ to $2.0$ in radius, and surface density
distribution $\Sigma \propto R^{-1}$, setup using the usual Monte-Carlo
technique. We also choose a vertical density profile corresponding to $H/R =
0.02$ at $R=1$ and employ an isothermal equation of state. To produce the two
distinct planes for the disc we rotate the outer half (in radius) of the disc
by an angle $\theta$. We then perform simulations with different values of
$\theta$ to find how the accretion rate changes with inclination angle.

\subsection{Tilted disc evolution}
\label{varytheta}
We perform seven simulations corresponding to inclination angles
$\theta=30n^{\circ}$ where $n=0,1,...,6$. The
$\theta=0^{\circ}$ simulation gives a flat--disc accretion rate to which we
compare the other simulations. We note that the $180^{\circ}$ case is
a rather unrealistic setup as it would require perfectly
anti-parallel accretion events. Even in the co-and counter-aligned
discs predicted by the numerical simulations of \citet{LP2006} the
discs never achieve a configuration where $\theta$ is precisely
$180^{\circ}$.

The simulations were run with an isothermal equation of state, i.e.
the temperature of each particle was held constant for each
simulation (see Section~\ref{adisims} below for a discussion of the
  effects of different equations of state).
This implies that the pressure, determined by
$P=c_{\rm s}^{2}\rho$, is simply proportional to $\rho$, and
the sound speed $c_{\rm s}$ is constant in time and the same
for all particles.

\begin{figure}
  \begin{center}
    \includegraphics[angle=0,width=\columnwidth]{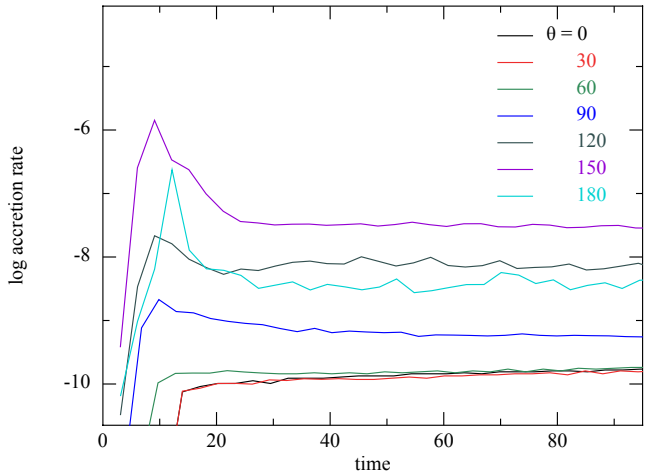}
    \caption{The accretion rates for the isothermal simulations with
      different inclination angles $\theta$. Time is in units of the
      dynamical time at $R=1$. The discs with a tilt $\ge 90^{\circ}$
      are much more efficient at driving accretion, as these involve a
      degree of counter--rotation and so direct cancellation of
      angular momentum.}
    \label{isoaccn}
  \end{center}
\end{figure}

In \fref{isoaccn} we report the accretion rates achieved for the
different values of $\theta$. This shows that for discs inclined by
less than $90^{\circ}$ the accretion rates are similar to that of a
flat disc for the duration of our simulations. In this case the
interaction between inclined gas orbits is relatively weak with the
gas staying on near circular orbits, with planes slowly varying from
the inner to outer disc. In contrast the simulations with $\theta \ge
90^{\circ}$ show a dramatic increase in the accretion rate. This is
caused by the direct cancellation of orbital angular momentum and
energy where the orbits of the disc gas cross.

\begin{figure}
  \begin{center}
    \includegraphics[angle=0,width=\columnwidth]{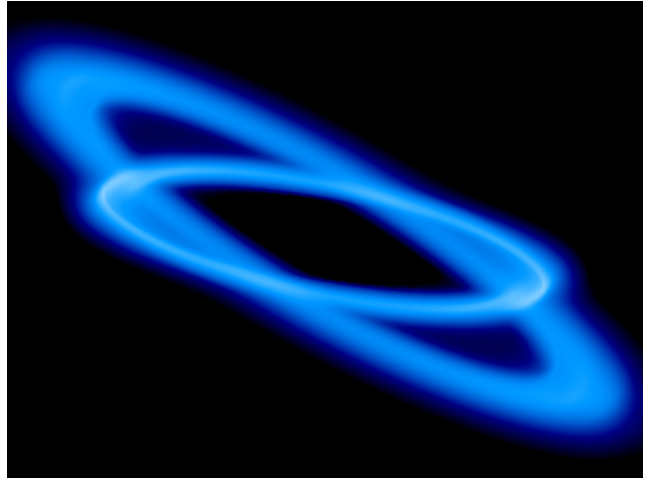}
    \caption{Column density projection showing the disc structure for the
      $\theta=30^{\circ}$ simulation. The disc has evolved for $\sim 50$
      dynamical times ($\sim 10$ orbits).}
    \label{theta30}
  \end{center}
\end{figure}

\begin{figure}
  \begin{center}
    \includegraphics[angle=0,width=\columnwidth]{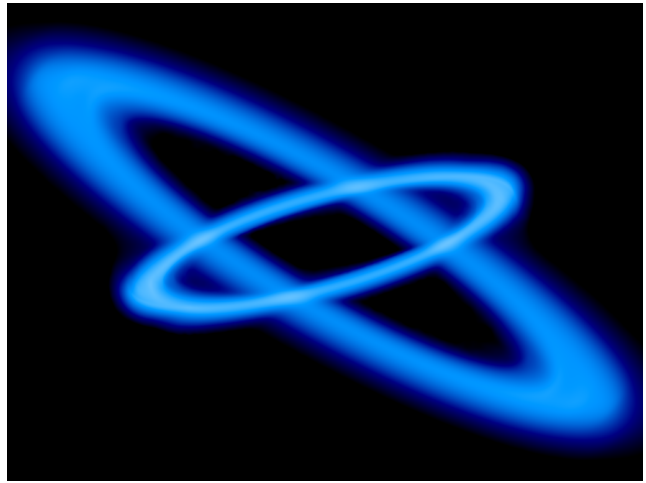}
    \caption{Column density projection showing the disc structure for the
      $\theta=60^{\circ}$ simulation. The disc has evolved for $\sim 50$
      dynamical times ($\sim 10$ orbits).}
    \label{theta60}
  \end{center}
\end{figure}

In \fref{theta30} we show the disc structure for the simulation where
$\theta=30^{\circ}$.
Here the two discs initially join to form a coherent disc with a
warped region joining the still misaligned inner and outer discs. Gas
is quickly depleted from the warp region, leaving a clear drop (of 2-3
orders of magnitude) in projected density in the warp region. This
break between the discs persists throughout the simulations as the two
discs slowly align with each other, reaching an angular separation
of $\sim 10^{\circ}$ by the time the simulation ends. The large
gradient in specific angular momentum in the warp region causes gas to
quickly move on past the warp. This causes the low density in the warp
region which is the disc break \citep{NK2012}. The disc
therefore maintains two distinct planes. We still resolve the rings of
gas between the two discs but this is clearly of much lower density
than the inner and outer discs (cf \citealt{LP2010}). In
\fref{theta60} we show the similar disc structure for the simulation
where $\theta=60^{\circ}$.

The $\theta = 90^{\circ}$ simulation (not illustrated) has an initial
period of chaotic flow between the two discs, until the inner disc
moves towards an inclination of $\sim 85^{\circ}$ after mixing with
some of the gas from the outer disc. At this point a strong warp is
set up and the flow continues in a similar way to $\theta=60^{\circ}$.

\begin{figure}
  \begin{center}
    \includegraphics[angle=0,width=\columnwidth]{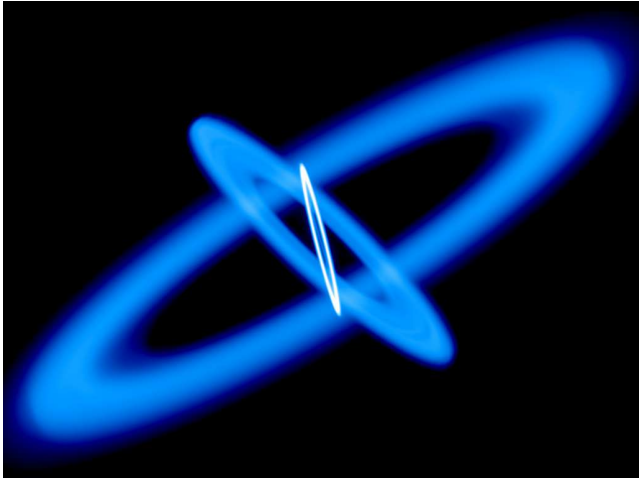}
    \caption{Column density projection showing the disc structure for
      the $\theta=120^{\circ}$ simulation. The disc has evolved for
      $\sim 50$ dynamical times ($\sim 10$ orbits). Note the new
      innermost disc (of high density and therefore white) which has
      formed from the dynamical infall of gas from the warp region.}
    \label{theta120}
  \end{center}
\end{figure}

The evolution of the $\theta=120^{\circ}$ and $\theta=150^{\circ}$
simulations proceed in a dramatically distinct way to those with
smaller $\theta$. This is simply because the sharing of angular
momentum now causes significant changes to the gas orbits.  In the
$\theta=120^{\circ}$ simulation gas falls from the region between the
two discs and circularises at a radius inside the inner accretion
disc. In this process some of gas is also accreted, as it carries a
spread of angular momentum. Clearly this gas does not have precisely
the angular momentum of the inner disc; it is a mix of gas from both
discs. Therefore the quasi--steady state for this simulation is an outer disc
causing cancellation with the outer edge of the original inner
disc. This gas falls and feeds both direct accretion and the new
innermost disc (white region of high density in \fref{theta120}). This
innermost disc is strongly warped with respect to the original inner
disc and so there is again extra dissipation between these two discs
as they try to straighten. The disc structure can be seen in
\fref{theta120}.

\begin{figure}
  \begin{center}
    \includegraphics[angle=0,width=\columnwidth]{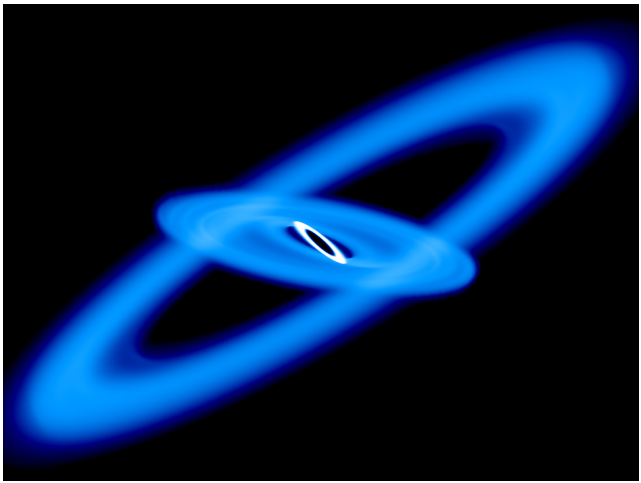}
    \caption{Column density projection showing the disc structure for the
      $\theta=150^{\circ}$ simulation. The disc has evolved for $\sim 50$
      dynamical times ($\sim 10$ orbits).}
    \label{theta150}
  \end{center}
\end{figure}

In the $\theta=150^{\circ}$ simulation the gas falls to a much smaller
radius from the region between the two discs. Initially a large amount
of gas is directly accreted on to the sink particle representing the
central accretor. The strong dissipation occurring early in the
simulation also causes the inner disc to spread all the way in to the
accretion radius. Some of the falling gas tries to circularise at a
radius where it is forced to impact upon the inner disc. As it still
has the sense of angular momentum of the outer disc this causes more
cancellation of angular momentum, driving more accretion. For
these reasons this simulation generates the highest accretion rates --
this is 
simply because it causes the greatest mixing of angular momentum in
the gas. The disc structure for this simulation can be seen in
\fref{theta150}.

\subsubsection{$\theta = 180^{\circ}$}
The $\theta=180^{\circ}$ case does not initially appear a very
realistic initial setup. However 
something close to this configuration may appear in the evolution of
accretion discs around spinning black holes (cf
\citealt{Kingetal2005}; \citealt{LP2006}) and also during the
evolution of an accretion disc around a binary system (cf
\citealt{Nixonetal2011b}). 

\begin{figure}
  \begin{center}
    \includegraphics[angle=0,width=\columnwidth]{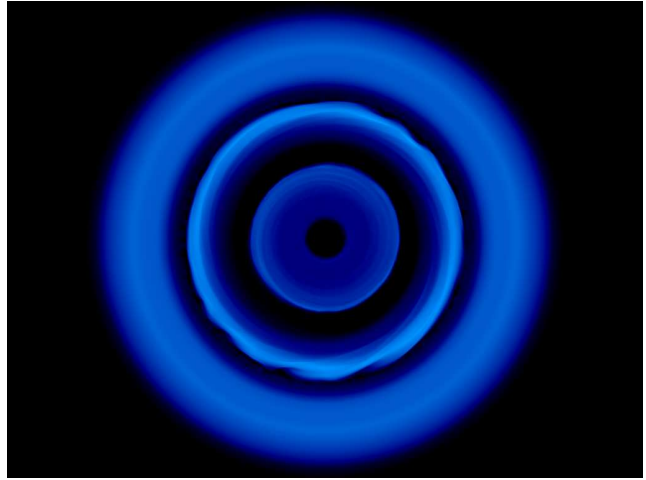}
    \caption{Column density projection showing the disc structure for
      the $\theta=180^{\circ}$ simulation. The disc has evolved for
      $\sim 50$ dynamical times ($\sim 10$ orbits). The innermost disc
      is formed of gas falling from the gap between the original inner
      disc (now the middle disc) and the outer disc. In this picture
      the innermost and outermost discs are rotating clockwise and the
      middle disc rotates counter--clockwise.}
    \label{theta180}
  \end{center}
\end{figure}

Initially while the two discs (inner and outer) are in contact there
is great cancellation of angular momentum for the gas. This gas then
attempts to fall towards the central sink particle. It therefore
impacts upon the inner disc. The pressure of the inner disc gas is
enough to force the falling gas out of the plane of the disc where it
can continue on a low angular momentum orbit towards the centre. It
then circularises inside the inner disc (with some accretion) after
colliding with gas on similar orbits and losing orbital energy in
shocks. This new innermost disc clearly has the same sense of angular
momentum as the outer disc since this has the larger specific angular
momentum. Once the original inner disc has spread inwards (and
this new innermost disc has spread outwards), exactly the same
process starts again between these two discs. See \fref{theta180} for
the disc structure.

Angular momentum cancellation thus proceeds as follows. Imagine the
two discs are not in contact. They spread viscously until gas from the
discs can interact, sharing and so destroying angular momentum. Now
consider two test particles (one from the inner disc and one from the
outer disc) which interact. The particle from the inner disc adopts a
slightly eccentric orbit, which passes through its parent disc. It
rejoins its own disc, which shrinks slightly to accommodate the loss
of angular momentum. The particle from the outer disc also adopts a
slightly eccentric orbit (inside its original orbit). But this forces
it to interact more with the `hostile' inner disc. This causes runaway
angular momentum loss for the outer disc particle, almost to the point
of free--fall to the centre. Now the particle's free--fall energy is
enough to move it out of the disc plane (helped by the pressure force
from the inner disc). It thus moves over and then inside the inner
disc. The overall effect is to shrink the inner disc and move the
outer disc particles on to orbits that pass inside the inner disc.

At this point one of three things can happen to the outer disc
particle: \\ 1) it may have cancelled enough angular momentum to
accrete directly. \\ 2) it may have cancelled enough to fall inside
the inner edge of the discs where it crosses the plane of the inner
disc and meets other particles on similar orbits: the particles shock
and dissipate energy, forming a new disc at much smaller radii. This
disc should be planar, but any small perturbations in the initial
discs will be amplified here at smaller radii and could produce a disc
with a substantial warp.  \\ 3) The particle's new orbit may take it
through the inner (initial) disc. If this is the case it again
interacts with the (hostile) disc and falls further. There are again
two possibilities here. Either the particle bounces all the way in
past the hostile disc, or it has so many interactions that it adopts
the sense of rotation of the hostile disc. So the particle either
accretes or drives the inner disc to shrink, and hence drives
accretion through the inner disc.

All sequences here greatly enhance the central accretion rate while
the discs are trying to mix their angular momenta.

\subsubsection{Effect of the gas equation of state}
\label{adisims}
\begin{figure}
  \begin{center}
    \includegraphics[angle=0,width=\columnwidth]{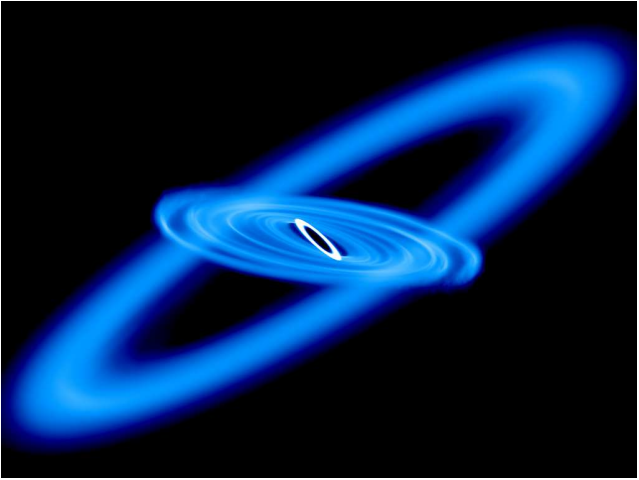}
    \caption{Column density projection showing the disc structure for the
      $\theta=150^{\circ}$ simulation with an isothermal equation of
      state. The disc has evolved for $\sim 20$ dynamical times.}
    \label{compiso}
  \end{center}
\end{figure}

\begin{figure}
  \begin{center}
    \includegraphics[angle=0,width=\columnwidth]{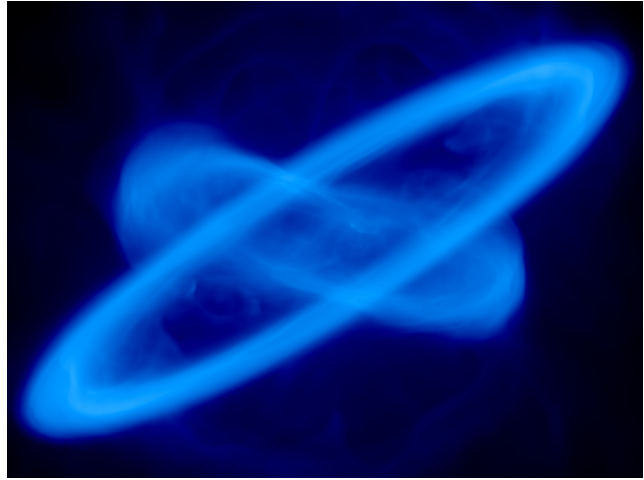}
    \caption{Column density projection showing the disc structure for the
      $\theta=150^{\circ}$ simulation with an ``adiabatic'' equation of state
      (see the text for a full definition). The disc has evolved for $\sim 20$
      dynamical times. Note that the gas no longer can cool to form a smaller
      disc (cf fig~\ref{compiso}), but instead forms outflows which return on
      chaotic orbits generating a lot of cancellation of angular momentum.}
    \label{compadi}
  \end{center}
\end{figure}

We have also investigated the effect of changing the thermal treatment of the
gas. In the simulations detailed above we used an isothermal equation of
state. The gas was assumed to radiate away any dissipative heating
instantly. The other extreme is to assume that any dissipation feeds the
internal energy of the particles, which are not allowed to cool by radiative
losses. Therefore we also simulate the $\theta=150^{\circ}$ case using an
adiabatic equation of state (with $\gamma=5/3$), transferring the kinetic
energy dissipated by the viscous terms into internal energy. In all other ways
this simulation is precisely the same as the corresponding isothermal
simulation. In Figs.~\ref{compiso} \& \ref{compadi} we show the disc evolution
after $\sim 20$ dynamical times. In the isothermal case the gas cancels
angular momentum and falls to circularise at a new radius. In comparison
Fig.~\ref{compadi} shows that the equation of state can play a significant
role in the hydrodynamics. In this case the gas is significantly heated when
it shocks and thus initially expands violently. This process drives more
mixing of the gas and thus the accretion is now almost a factor of two higher
here than in the isothermal case. The gas is still bound to the black hole and
so returns on chaotic orbits to eventually accrete.

\section{Discussion}
\label{discussion}
Retrograde circumbinary accretion discs are natural in the chaotic
accretion picture (\citealt{KP2006}; \citealt{KP2007}), and offer
the possibility of cancelling the binary orbital angular momentum
\citep{Nixonetal2011a}.  We have performed numerical simulations of
accretion discs with two initially distinct planes. In the region
between the two discs angular momentum is shared, driving accretion,
particularly if $\theta > \pi/2$.

The highest accretion rates come from discs with inclination $\theta
\approx 150^{\circ}$. Here the gas cancels angular momentum when the
discs meet, allowing this gas to fall directly to the centre without
the need to interact with the inner disc. Clearly this gas would in
reality have some residual angular momentum and so would circularise
at a new smaller radius. In the simulations however the gas falls
inside our accretion radius for the central sink particle and so is
directly accreted. Gas falling from between the discs always has
angular momentum with the same sense of direction as the outer disc.
Therefore if the gas tries to circularise outside the accretion radius
it impacts the inner disc, cancelling more angular momentum and
falling further until it is accreted.

The accretion rates found here are $>100$ times those of a flat planar
disc. This increase comes from the mixing of angular momenta and consequent
dynamical infall of gas to smaller radii. As gas is depleted from the gap the
disc tries to spread and fill the gap, promoting more cancellation of angular
momentum and so more accretion. The timescale for gas accretion is shortened
(cf. equation~\ref{tvisc}). If enough angular momentum is cancelled then the
gas accretes directly on a dynamical timescale.

Our simulations show that discs with an initial inclination less than
$90^{\circ}$ evolve into coherent discs joined by a warp. Over time,
dissipation from this warp brings the two discs into a common plane. This
extra dissipation enhances accretion, the effect increasing with inclination
angle (although negligible in comparison to the enhanced accretion when
$\theta>\pi/2$).  The discs appear to break, in the sense that the surface
density of the disc is greatly reduced in the warp. However in these
high--resolution simulations we still resolve the orbits of the disc particles
in the warp. This disc breaking is predicted by the numerical simulations of
warp propagation in accretion discs by \citet{NK2012} using the constrained
viscosities of \citet{Ogilvie1999}, and has been also been seen in the SPH
simulations of \citet{LP1997} and \citet{LP2010}.

Inclinations greater than $90^{\circ}$ cancel rather than sharing
angular momentum between particles. This leads to particles falling on
to smaller orbits, and so the accretion rates in these simulations are
significantly higher. The accretion rate for $\theta=180^{\circ}$ is
lower than that of $\theta=150^{\circ}$. This is probably because gas
must be pushed out of the plane of the discs before it can leap over
the inner disc.

The numerical simulations do not predict accretion rates as high as
the analytical approach (\ref{rcirc1}). This was based on two
assumptions: that the angular momentum cancellation is perfect, and an
equal mass from each disc interacts. The simulations suggest instead
that the gas falling from the gap has a spread of angular momentum, where
some accretes and some falls to a smaller radius.

The main result of this paper is that internally tilted accretion discs can
generate enhanced accretion rates of up to $\sim 10^{4}$ times that of a plane
disc (cf \fref{isoaccn}). In a quasi--steady state (such discs never reach a
full steady state) we see accretion rates that are $\gtrsim 100$ times those
of a planar disc.  As remarked above, discs tilted in this way arise naturally
during the alignment of a disc around a spinning black hole or a central
binary (cf. \citealt{LP2006} \& \citealt{Nixonetal2011b}). We also expect such
structures to occur during chaotic gas infall into galactic centres. This is a
very efficient way of driving gas right down to the vicinity of an SMBH or an
SMBH binary and thus causing accretion. Nested discs suggest a way of
transferring gas from large radii (approaching galaxy scales) to small radii.

We note finally that our result, that tilted and nested discs imply
enhanced accretion, has implications for simulations where subgrid
models are used to mimic feedback from the SMBH. The dynamics of
misaligned accretion events may imply much larger accretion rates
than predicted by the simple assumption of a planar disc with a fixed
direction of angular momentum.
\section*{Acknowledgments}
\label{acknowledgements}
We thank the referee for useful feedback. We acknowledge the use of
\textsc{splash} \citep{Price2007} for the rendering of the SPH plots. CJN is
supported by an STFC studentship. Theoretical astrophysics research at
Leicester is supported by an STFC rolling grant. This research used the ALICE
High Performance Computing Facility at the University of Leicester.  Some
resources on ALICE form part of the DiRAC Facility jointly funded by STFC and
the Large Facilities Capital Fund of BIS.

\bibliographystyle{mn2e} 
\bibliography{nixon}

\begin{thebibliography}{}

\bibitem[\protect\citeauthoryear{{Fanidakis}, {Baugh}, {Benson}, {Bower},
  {Cole}, {Done} \& {Frenk}}{{Fanidakis} et~al.}{2011}]{Fanidakisetal2011}
{Fanidakis} N.,  {Baugh} C.~M.,  {Benson} A.~J.,  {Bower} R.~G.,  {Cole} S.,
  {Done} C.,    {Frenk} C.~S.,  2011, \mnras, 410, 53

\bibitem[\protect\citeauthoryear{{King}, {Lubow}, {Ogilvie} \&
  {Pringle}}{{King} et~al.}{2005}]{Kingetal2005}
{King} A.~R.,  {Lubow} S.~H.,  {Ogilvie} G.~I.,    {Pringle} J.~E.,  2005,
  \mnras, 363, 49

\bibitem[\protect\citeauthoryear{{King} \& {Pringle}}{{King} \&
  {Pringle}}{2006}]{KP2006}
{King} A.~R.,  {Pringle} J.~E.,  2006, \mnras, 373, L90

\bibitem[\protect\citeauthoryear{{King} \& {Pringle}}{{King} \&
  {Pringle}}{2007}]{KP2007}
{King} A.~R.,  {Pringle} J.~E.,  2007, \mnras, 377, L25

\bibitem[\protect\citeauthoryear{{King}, {Pringle} \& {Hofmann}}{{King}
  et~al.}{2008}]{Kingetal2008}
{King} A.~R.,  {Pringle} J.~E.,    {Hofmann} J.~A.,  2008, \mnras, 385, 1621

\bibitem[\protect\citeauthoryear{{Larwood} \& {Papaloizou}}{{Larwood} \&
  {Papaloizou}}{1997}]{LP1997}
{Larwood} J.~D.,  {Papaloizou} J.~C.~B.,  1997, \mnras, 285, 288

\bibitem[\protect\citeauthoryear{{Lodato} \& {Price}}{{Lodato} \&
  {Price}}{2010}]{LP2010}
{Lodato} G.,  {Price} D.~J.,  2010, \mnras, 405, 1212

\bibitem[\protect\citeauthoryear{{Lodato} \& {Pringle}}{{Lodato} \&
  {Pringle}}{2006}]{LP2006}
{Lodato} G.,  {Pringle} J.~E.,  2006, \mnras, 368, 1196

\bibitem[\protect\citeauthoryear{{Murray}}{{Murray}}{1996}]{Murray1996}
{Murray} J.~R.,  1996, \mnras, 279, 402

\bibitem[\protect\citeauthoryear{{Nixon}, {Cossins}, {King} \&
  {Pringle}}{{Nixon} et~al.}{2011}]{Nixonetal2011a}
{Nixon} C.~J.,  {Cossins} P.~J.,  {King} A.~R.,    {Pringle} J.~E.,  2011,
  \mnras, 412, 1591

\bibitem[\protect\citeauthoryear{{Nixon} \& {King}}{{Nixon} \&
  {King}}{2012}]{NK2012}
{Nixon} C.~J.,  {King} A.~R.,  2012, \mnras, p.~2449

\bibitem[\protect\citeauthoryear{{Nixon}, {King} \& {Pringle}}{{Nixon}
  et~al.}{2011}]{Nixonetal2011b}
{Nixon} C.~J.,  {King} A.~R.,    {Pringle} J.~E.,  2011, \mnras, 417, L66

\bibitem[\protect\citeauthoryear{{Ogilvie}}{{Ogilvie}}{1999}]{Ogilvie1999}
{Ogilvie} G.~I.,  1999, \mnras, 304, 557

\bibitem[\protect\citeauthoryear{{Price}}{{Price}}{2007}]{Price2007}
{Price} D.~J.,  2007, \pasa, 24, 159

\bibitem[\protect\citeauthoryear{{Shakura} \& {Sunyaev}}{{Shakura} \&
  {Sunyaev}}{1973}]{SS1973}
{Shakura} N.~I.,  {Sunyaev} R.~A.,  1973, \aap, 24, 337

\end{thebibliography}

\end{document}